# Measurement of infrared optical constants with visible photons

Anna Paterova[1,2], Hongzhi Yang[1], Chengwu An[1], Dmitry Kalashnikov[1], and Leonid Krivitsky[1,*]

[1] *Data Storage Institute, Agency for Science Technology and Research (A\*STAR), 138634 Singapore*

[2] *School of Electrical and Electronic Engineering, Nanyang Technological University, 639798 Singapore*

\*E-mail: Leonid-K@dsi.a-star.edu.sg

**We demonstrate a new scheme for infrared spectroscopy with visible light sources and detectors. The technique relies on the nonlinear interference of correlated photons, produced via spontaneous parametric down conversion in a nonlinear crystal. Visible and infrared photons are split into two paths and the infrared photons interact with the sample under study. The photons are reflected back to the crystal, resembling a conventional Michelson interferometer. Interference of the visible photons is observed and it is dependent on the phases of all three interacting photons: pump, visible and infrared. The transmission coefficient and the refractive index of the sample in the infrared range can be inferred from the interference pattern of visible photons. The method does not require the use of potentially expensive and inefficient infrared detectors and sources, it can be applied to a broad variety of samples, and it does not require *a priori* knowledge of sample properties in the visible range.**

## I. INTRODUCTION

Infrared (IR) spectroscopy is a powerful tool for numerous applications including material analysis, environmental sensing, health diagnostics and others. Fourier Transform Infrared (FTIR) spectroscopy is a dominant technology in this field [1]. FTIR spectrometers are available as bench top and hand held instruments, and they yield broad operation range, high spectral resolution, and fast readout. However, these instruments face challenges associated with high cost and non-ideal performance of IR-range light sources, optical components, and photon detectors.

The phenomena of nonlinear and quantum optics allow circumventing the need for IR-range components for IR spectroscopy. For the case of up-conversion spectroscopy, two visible range lasers (one with a fixed frequency, and one with a tunable frequency) are mixed in a nonlinear crystal to produce a difference frequency signal in the IR range [2, 3]. The IR signal is then used to probe the transmission of a sample and is converted back to the visible range using a frequency up-conversion crystal. One major disadvantage of this method is the requirement of sophisticated tunable lasers, optical system and multichannel detection.

Another approach uses the effect of nonlinear interference of photons produced via spontaneous parametric down conversion (SPDC) [4-10]. In SPDC, a photon from the pump laser decays into a pair of correlated photons in a nonlinear medium. The nonlinear interferometer consists of two nonlinear crystals, arranged in such a way that down-converted photons from both crystals are superimposed. Although the photons may have different wavelengths, the interference pattern of the signal photons depends on the phase and transmissivity of the idler photons. Mandel and colleagues discovered this effect, which is now known as induced coherence [4, 5]. Recently several works have demonstrated the application of this technique for imaging [6], metrology [7] and spectroscopy [8, 9, 11-13]. This approach benefits from much simpler realization in comparison with up-conversion spectroscopy [2, 3], because it uses vacuum fluctuations of the field instead of tunable lasers. Another feature of the technique is that the same setup can be used for measurements of the absorption coefficient and the refractive index in the IR range. This feature is distinctively advantageous over conventional methods, which require the use of sophisticated IR-ellipsometers combined with FTIR spectrometers.

In the earlier nonlinear spectroscopy schemes two nonlinear crystals were set sequentially into a common pump beam [8, 9]. A sample was placed between the crystals and all the three photons (pump, signal, and idler) passed through it. The absorption spectrum of the sample in the IR was inferred from the interference of visible photons. The method does not require the use of IR equipment, and benefits in high stability of interferometric scheme due to degenerate arms of the interferometer. However, it has several practical limitations. The major one is that the sample has to be transparent to both signal and pump photons, and sample's optical properties at these wavelengths have to be known in advance. The sample also introduces additional scattering of the pump, which may also damage it.

In this work, we introduce an alternative approach to IR spectroscopy with visible light, which is free of the limitations mentioned above. We build a setup in the form of a nonlinear Michelson interferometer, where signal (in the visible range) and idler (in the IR range) photons are split using a dichroic beam splitter and then reflected back into the crystal by two mirrors, see Fig. 1. The sample under test is placed into the path of IR photons, and it modifies the interference pattern for visible photons. We infer the transmission coefficient of the sample and its refractive index in the IR-range from the modified interference pattern of visible photons. Although the alignment of this scheme is more challenging in comparison with the one reported in [8 ,9], it is appealing from the practical standpoint. Unique benefits of the current method compared to the one reported in Refs.[8, 9] are: (1) it works for samples opaque in the visible range; (2) it does not require apriori knowledge of sample properties in the visible range; (3) it eliminates damage to the sample by the pump; (4) it prevents scattering of the pump and signal photons. The work further adds to the number of optical characterization techniques, which benefit from quantum enhancement effects [14-19].

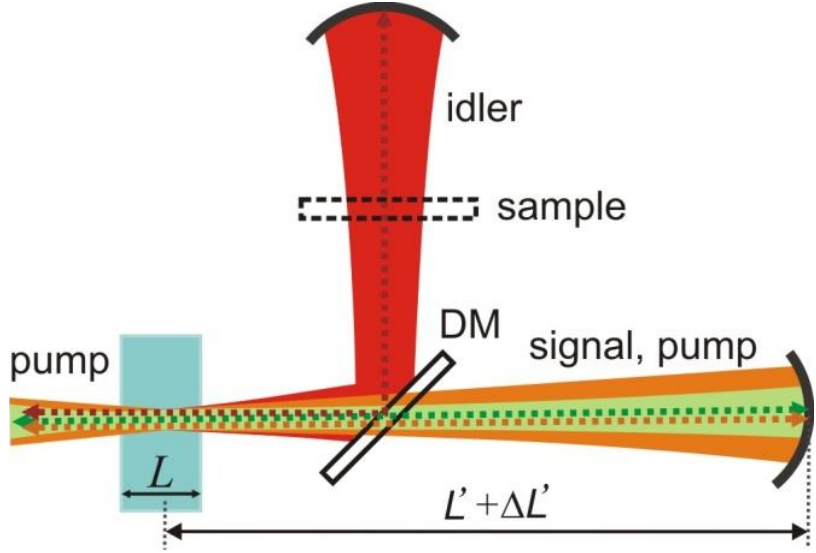

FIG. 1. A nonlinear version of a Michelson interferometer. Signal and idler photons are generated via SPDC in a nonlinear crystal. The idler photon (IR) is separated from the signal and pump photons (visible) by a dichroic mirror (DM). Spherical mirrors are positioned at a distance of $L' = 2F'$ from the crystal, where $F'$ is the focal lengths of the mirrors. The sample is inserted in the path of the idler photon. $\Delta L'$ is the optical path lengths difference between the arms. The interference pattern of signal photons is detected at the output of the interferometer.

## II. MATERIALS AND METHODS

### A. Theory

Let us consider the experimental scheme in Fig. 1, which represents a nonlinear version of the Michelson interferometer. At the first pass through a nonlinear crystal, the pump creates a pair of correlated photons in the visible (signal) and IR (idler) range via SPDC. Next, down-converted photons are split using a dichroic mirror and reflected back to the crystal by the mirrors. Pump and signal photons propagate together. The reflected pump beam generates down-converted photons at the second pass. A medium of interest is inserted into the path of idler photons.

Due to energy conservation and phase matching requirements, the frequencies $\omega$ and the wave vectors $\vec{k}$ of the down-converted photons are given by [20]:

$$\omega_p = \omega_s + \omega_i \, ; \, \vec{k}_p = \vec{k}_s + \vec{k}_i + \Delta\vec{k}, \qquad (1)$$

where sub-indices $p, s, i$ correspond to pump, signal and idler photons respectively, and $\Delta\vec{k}$ is a wave vector mismatch. In the first-order approximation of the perturbation theory the state vector of the down-converted filed produced in a crystal can be written as [11, 21]:

$$|\psi\rangle = |vac\rangle + \eta \sum_{k_s, k_i} F(\vec{k}_s, \vec{k}_i) a^+_{k_s} a^+_{k_i} |vac\rangle, \qquad (2)$$

where $\eta \ll 1$ is the SPDC conversion factor of the crystal; summation indices $k_s, k_i$ indicate frequency, angular and polarization modes of signal and idler photons; $a^+_{k_s}$, $a^+_{k_i}$ are photon creation operators for signal and idler photons respectively; $F(\vec{k}_s, \vec{k}_i)$ is the two-photon field amplitude, which is defined by the following function [11, 21]:

$$F(\vec{k}_s, \vec{k}_i) \propto \int_V E_p(\vec{r}) \chi(\vec{r}) \exp\left[i\Delta k(\vec{k}_s, \vec{k}_i)\vec{r}\right] d^3\vec{r}, \qquad (3)$$

where $E_p(\vec{r})$ is the amplitude of the pump, $\chi(\vec{r})$ is the nonlinear susceptibility of the crystal. The integral in Eq. (3) is taken over the interaction volume in the crystal $v$. In the approximation of a monochromatic plane wave pump, collinear phase matching, a uniform ($\chi(\vec{r}) = const$), infinite (in the transverse direction) and thin (in the longitudinal direction) crystal Eq. (3) takes the form [11, 22, 23]:

$$F(\vec{k}_s, \vec{k}_i) \propto E_p(k_{s\perp}, k_{i\perp}) \int_0^L \exp\left[i\Delta k_\| z\right] dz \propto E_p(k_{s\perp}, k_{i\perp}) \operatorname{sinc}\left(\frac{\Delta k_\| L}{2}\right), \qquad (4)$$

where $L$ is the length of the crystal, $z$ is the axis along the pump beam, $\Delta k_\|$ is the longitudinal component of the wave vector mismatch parallel to the $z$ axis, $k_{s\perp}$ and $k_{i\perp}$ are transverse components of wave vectors, perpendicular to $z$ axis. Function $F(\vec{k}_s, \vec{k}_i)$ defines the angular-frequency spectrum of the two-photon field. Its angular spectrum is determined by the interplay between the divergence of the pump beam and the phase-matching conditions in the crystal [22, 23].

Here we consider the case of a single spatial mode of the SPDC and limit our analysis to spectral characteristics of the two-photon amplitude. We can rewrite the two-photon amplitude $F(\vec{k}_s, \vec{k}_i)$ in terms of frequencies $\omega_s$ and $\omega_i$.

The resulting state vector in the interferometer is given by a superposition of state vectors of down-converted photons generated from the first (1) and the second (2) pass of the pump through a nonlinear crystal:

$$|\Psi\rangle = |\psi\rangle_1 + |\psi\rangle_2 = |vac\rangle + \tilde{\eta} \iint d\omega_s d\omega_i F(\omega_s, \omega_i) a_{s1}^+ a_{i1}^+ |vac\rangle + \\ + \tilde{\eta} e^{i\varphi_p} \iint d\omega_s d\omega_i F(\omega_s, \omega_i) a_{s2}^+ a_{i2}^+ |vac\rangle, \qquad (5)$$

where $s1,2$ and $i1,2$ indicate spatial modes of down-converted photons with frequencies $\omega_s$ and $\omega_i$, $\varphi_p$ is the phase acquired by the pump [24- 26]. Here we assume that the frequency mode spacing is small and substitute the summation over frequency modes by the integration [5, 10, 27]. We also assume that losses introduced by optical elements in the setup are equal for signal and pump photons. Therefore, interference terms in the superposition state vector, given by Eq. (5), have equal weights $\tilde{\eta}$. Parameter $\tilde{\eta}$ accounts for the conversion efficiency and losses of pump and signal photons in the interferometer.

Losses induced by the medium in the idler arm can be described by a beam splitter model, in which a vacuum state is introduced into the empty port of the beam splitter [28]. Then, the photon creation operator for the idler photons at the second pass of the pump is given by [4, 5, 29]:

$$a_{i2} = e^{i\tilde{\varphi}_i}\left(\tau^2 a_{i1} + r\tau a_{0'} + r a_{0''}\right), \tag{6}$$

where $a_{0'}$, $a_{0''}$ describe vacuum fields entering from unused ports of the beam splitter with frequency $\omega_i$, $\tau$ and $r$ are complex amplitude transmission and reflection coefficients of idler photons for a single pass through the medium, $\tilde{\varphi}_i$ is the phase acquired by the photon in mode $i1$. Then, the resulting state vector (5) is given by:

$$|\Psi\rangle = |vac\rangle + \tilde{\eta}\iint d\omega_s d\omega_i F(\omega_s,\omega_i) a_{s1}^+ a_{i1}^+ |vac\rangle + \\ + \tilde{\eta}\iint d\omega_s d\omega_i F(\omega_s,\omega_i) a_{s2}^+ e^{i(\varphi_p - \tilde{\varphi}_i)}\left(\tau^{*2} a_{i1}^+ + r^*\tau^* a_{0'}^+ + r^* a_{0''}^+\right)|vac\rangle \tag{7}$$

The count rate of signal photons is given by: $P_s \propto \langle\Psi|E_s^{(-)}E_s^{(+)}|\Psi\rangle$, where $E_s^{(+)}(t) \propto \int d\omega\left(a_{s1}(\omega)e^{-i\omega(t-t_1)} + a_{s2}(\omega)e^{-i\omega(t-t_2)}\right)$ is the electric field at the detector, and $t_{1,2}$ are travel times of signal photons from the crystal to the detector. Using Eq. (7) and taking into account that $\tilde{\varphi}_i = \omega_i t_0$, where $t_0$ is the travel time of the idler photons in the interferometer, we obtain the following equation:

$$E^{(+)}|\Psi\rangle \propto \iint d\omega_s d\omega_i F(\omega_s,\omega_i)\left[a_{i1}^+ e^{-i\omega_s(t-t_1)} + \\ + e^{-i\omega_s(t-t_2)+i(\varphi_p - \omega_i t_0)}\left(\tau^{*2} a_{i1}^+ + r^*\tau^* a_{0'}^+ + r^* a_{0''}^+\right)\right]|vac\rangle \tag{8}$$

Then the count rate is:

$$P_s \propto \iint d\omega_s d\omega_i |F(\omega_s,\omega_i)|^2 \left[\left|e^{-i\omega_s(t-t_1)} + \tau^{*2}e^{-i\omega_s(t-t_2)+i(\varphi_p - \omega_i t_0)}\right|^2 + \\ + \left|r^*\tau^* e^{-i\omega_s(t-t_2)+i(\varphi_p - \omega_i t_0)}\right|^2 + \left|r^* e^{-i\omega_s(t-t_2)+i(\varphi_p - \omega_i t_0)}\right|^2\right], \tag{9}$$

Redefining frequencies as $\omega_s = \omega_{0s} - \Omega$, $\omega_i = \omega_{0i} + \Omega$, where $\omega_{0s,0i}$ are central frequencies of signal and idler photons, and $\Omega$ is the detuning frequency, the photon count rate is:

$$P_s \propto \int d\Omega |F(\Omega)|^2 \left(|\tau|^2|r|^2 + |r|^2 + \left|e^{-i(\omega_{s0}-\Omega)(t-t_1)} + \tau^{*2}e^{-i((\omega_{s0}-\Omega)(t-t_2)-\varphi_p+(\omega_{i0}+\Omega)t_0)}\right|^2\right), \tag{10}$$

Taking into account the normalization of the two-photon amplitude $\int|F(\Omega)|^2 d\Omega = 1$, Eq. (10) takes the form:

$$P_s \propto 2 + \left(\tau^{*2} e^{i(\varphi_p + \omega_{s0}(t_2-t_1)-\omega_{i0}t_0)}\int|F(\Omega)|^2 e^{-i\Omega(t_2+t_0-t_1)}d\Omega + c.c.\right), \tag{11}$$

where *c.c.* stands for complex conjugate. Assuming that $\omega_{s0}(t_1 - t_2) = \varphi_s$ and $\omega_{i0}t_0 = \varphi_i$ are accumulated phases of signal and idler photons in the interferometer, the final expression for the count rate at the detector has the following form:

$$P_s \propto 2\left[1+|\tau_i|^2 |\mu(\Delta t)|\cos\left(\varphi_p - \varphi_s - \varphi_i + \arg\tau^2 + \arg\mu(\Delta t)\right)\right], \tag{12a}$$

$$\mu(\Delta t) = \int |F(\Omega)|^2 e^{-i\Omega\Delta t} d\Omega, \tag{12b}$$

where $\mu(\Delta t)$ is the normalized correlation function of the SPDC, $\mu(0) = 1$ [5, 30], and $\Delta t = t_0 + t_2 - t_1$ is the time delay. The delay $\Delta t$ is equal to zero when the optical path lengths of signal and idler photons in the interferometer are the same, see Fig. 1:

$$n_s L + n'_s L'_s = n_i L + n'_i L'_i, \tag{13}$$

where $n_{s,i}$ are the refractive indices of the nonlinear crystal at signal and idler wavelengths, $n'_{s,i}$ are the refractive indices of the medium at signal and idler wavelengths, and $L'_{s,i}$ are the path lengths for signal and idler photons in the interferometer.

If the phases of the signal and of the pump photons are fixed, then the variation of the photon count rate $P_s$ is determined by the phase of the idler photons $\Delta\varphi_i = 2\frac{2\pi n'_i}{\lambda_i}\Delta x$, where $\lambda_i$ is the idler photon wavelength, and $\Delta x$ is the displacement of the IR mirror. Then the visibility of the interference fringes is given by:

$$V = \frac{P_s^{max} - P_s^{min}}{P_s^{max} + P_s^{min}} = |\mu(\Delta t)||\tau|^2. \tag{14}$$

From Eq. (14) it follows that the transmission coefficient of the idler (IR) photons through the medium can be inferred from the visibility of interference of signal photons detected in the visible range.

The experimental procedure for measuring the transmission spectra of the medium is as follows. First, we equalize the interferometer arms and measure the interference with the air as a reference medium $n'_{s,i} = n^{air}_{s,i} = 1$. From the measurements of the interference visibility $V_{ref}$ we infer the total intrinsic losses of our system $|\tau_{app}|^2$, see Eq. (14). Then we insert the medium of interest. We align the interferometer arms to compensate the incurred delay:

$$\Delta L' = 2\left(n'_i - n^{air}_i\right)L_m, \tag{15}$$

where $L_m$ is the thickness of the medium. Then we measure the interference visibility again and infer its relative change, see Eq. (10):

$$\frac{V}{V_{ref}} = \frac{|\tau_{app}|^2 |\tau|^2}{|\tau_{app}|^2} = |\tau|^2 \equiv T, \tag{16}$$

We infer the refractive index and the transmission coefficient of the medium in the IR from the change of the optical path, and the visibility, respectively, see Eqs. (15), (16). From the obtained values, we infer the absorption $\alpha$ and reflection $R$ coefficients of the medium at the corresponding wavelength:

$$V \propto |\tau|^2 = \left( \left(1-|r|^2\right) e^{-\frac{\alpha L_m}{2}} \right)^2 = (1-R)^2 e^{-\alpha L_m}, \quad (17a)$$

$$|r|^2 \equiv R = \left( \frac{n'_i - n_i^{air}}{n'_i + n_i^{air}} \right)^2, \quad (17b)$$

where $R$ is reflection coefficient calculated at normal incidence.

### B. Experiment

Our experimental setup is shown in Fig. 2. We use a continuous wave laser at 532 nm as a pump (Laser Quantum Torus, power 150 mW, coherence length ~100 meters). A dichroic mirror $DM_1$ (Semrock) reflects the pump beam to the nonlinear interferometer and transmits the signal photons in the range of 562-950 nm. The nonlinear Michelson interferometer consists of a MgO:LiNbO$_3$ crystal (L=0.5 mm; Dayoptics), a gold-coated dichroic mirror $DM_2$ (ISP Optics), and two spherical mirrors (f=50 mm, Thorlabs): one is silver coated ($M_s$), and another one is gold coated ($M_i$).

The pump beam is focused into the crystal by the lens $F_1$ (f=200 mm). Signal (visible) and idler (IR) photons are generated via type-I (e → oo) collinear frequency non-degenerate SPDC. The tuning of the central wavelength of the down-converted photons is performed by tilting the crystal. The idler photon wavelength is inferred from the wavelength of signal photons according to Eq. (1). The dichroic mirror $DM_2$ reflects the idler photons and transmits the pump and signal photons.

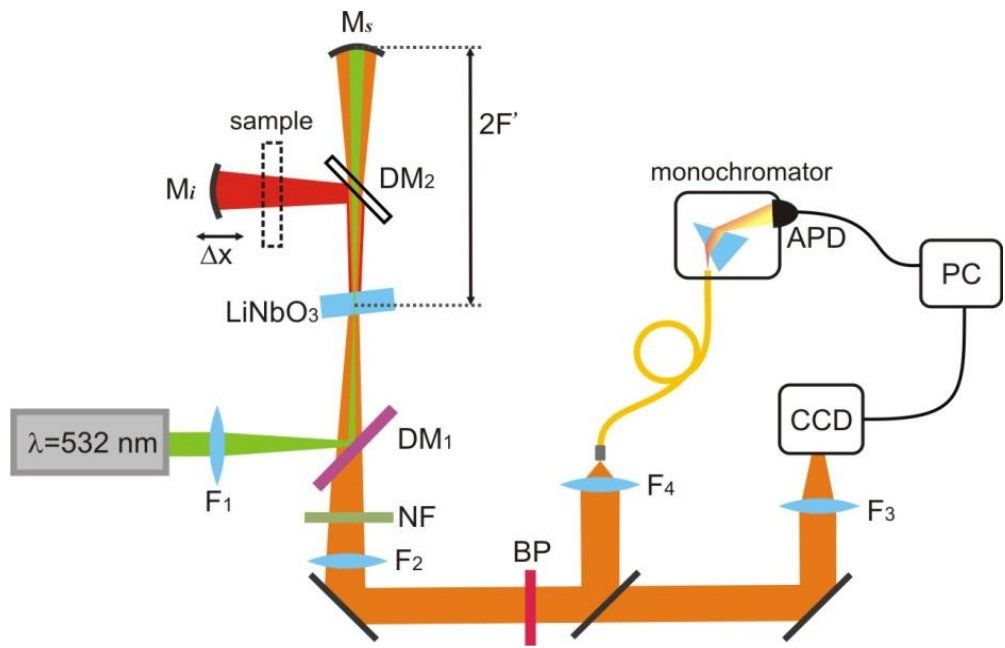

FIG. 2. The experimental setup. A CW-laser at 532 nm is used as a pump for a LiNbO$_3$ crystal, where SPDC photons are generated. Idler photons are reflected by the dichroic mirror $DM_2$, while signal and pump

photons pass through the mirror. Spherical mirrors $M_s$ and $M_i$ with focal distance $F'$ reflect the photons back into the crystal. A sample is placed in the IR arm of the interferometer between $DM_2$ and $M_i$. The interference pattern is captured either by a CCD camera or by a single photon detector (APD) preceded by a monochromator. NF is a notch filter; BP is a bandpass filter.

The visibility of the interference depends on the spatial overlap of the interfering photons [6, 31]. The intrinsic divergence of the SPDC results in different phases acquired by transverse components of wave vectors, which reveals itself in a pronounced ring structure [9, 32]. To circumvent this problem, we focus the pump beam inside the crystal, so that the divergence of down-converted photons is defined by the divergence of the pump beam. We use spherical mirrors $M_{s,i}$, positioned at a distance of their curvature radius from the crystal to form a one-to-one image of the pump waist in the crystal. Use of metallic spherical mirrors eliminates the problem of the chromatic dispersion. In this way, SPDC beams produced in two paths can be perfectly overlapped.

In the detection part, signal photons are separated from the pump by the dichroic mirror $DM_1$ and the notch filter NF (Semrock). The signal photons are then collimated by a lens $F_2$ (f=200 mm). The interference pattern can be detected in two ways:

1. Beams of signal photons are detected using a silicon CCD camera (Andor, iXon3 897). The CCD is located at the focal plane of the lens $F_3$ (f=500 mm), and it images the beam of signal photons. Measurements with the CCD camera facilitate alignment of the setup.
2. Signal photons emitted collinearly to the pump beam are detected by a single photon avalanche photodiode (APD, Perkin Elmer) preceded by a home-built monochromator. The monochromator allows performing wavelength selective measurements (resolution 0.25 nm) at different wavelengths of signal photons.

For the coarse alignment of the interferometer, the mirror $M_i$ is translated using a motorized translation stage until a clear interference pattern is observed on the CCD. Then, signal photons are directed to the APD bypassing the monochromator. The dependence of the intensity on the optical path difference with and without the sample is measured by translating the mirror $M_i$. Using a laser with a long coherence length ensures that the change of the interference pattern with the displacement of the mirror is defined only by the coherence length of SPDC photons. From the shift of the maxima of the interference pattern, we infer the refractive index of the sample in the IR range according to Eq. (15).

Then, we perform a fine scan of the interference fringes with and without the sample. Signal photons are spectrally filtered by the monochromator and detected by the APD. We translate the mirror $M_i$ using a piezo actuator (Thorlabs) and measure visibility of the fringes. We infer the transmission coefficient of the sample at the wavelength of idler photon according to Eq. (16).

We measured that the phase drift in the interferometer was slower than the typical measurement time of one fringe (~30 sec). It is certainly possible to implement active stabilization of the interferometer, for instance, by monitoring the intensity of the residual pump beam exiting from the empty port of $DM_2$.

## III. RESULTS AND DISCUSSION

We apply our technique to study four different samples: polydimethylsiloxane (PDMS), BK7 glass and pure silicon windows with and without anti-reflection (AR) coating (Edmund Optics). For proof-of-concept demonstration of our technique we chose the spectral range within 2100-2900 nm, which corresponds to the observation of signal photons at the range of 710-655nm.

First, we measure the interference pattern without the sample. The dependence of APD photocounts on the position of the mirror $M_i$ is shown in Fig. 3 (black squares). We fix the position at the point, where the visibility is maximized, and proceed with the fine scan using the piezo actuator. Black squares show the interference fringe detected by the APD, see inset in Fig. 3. Note that the period of the fringes corresponds to the wavelength of idler photons, which agrees with the theoretical prediction, see Eq. (12a). Without a sample, the visibility is about 14.00±0.38%. The main factor, which limits the visibility of the interference is the imperfection of the optical elements. From our independent measurements of the $DM_2$, we found that for the IR light (2100-2900 nm) the reflection is 20% and for the visible light (650-750 nm) the transmission is 70%. The nonlinear crystal has the absorption of about 20%. From Eq. (14) we estimate that the expected visibility is about 15%, which is in agreement with our experiment. Note that further improvements in the visibility are possible by optimization of coatings and materials of optical elements.

The envelope function of the interference signal in Fig.3 is given by the envelope of the first order correlation function of the SPDC. Since the SPDC spectrum has a $sinc^2$ form (see Eq.4), the correlation function has a triangular shape, with amplitude given by the visibility of the interference and the width given by the coherence length of the SPDC. In Fig.3 we plot the theoretical function with parameters obtained from experimentally measured SPDC spectrum (width at $\lambda_s$=694 nm is $\Delta\lambda_s$=5.8 nm; corresponding coherence length is 83 μm) and visibility values.

Next, we insert a sample (PDMS, ~143 μm thick) in the IR beam. The interference pattern shifts and its visibility is reduced, see Fig. 3 (red circles). We translate the mirror $M_i$ for approximately 120 μm to recover the interference. Then we select the wavelength by the monochromator and perform a fine scan of the interference fringes, see inset in Fig. 3 (red circles). From the change of optical path length, we infer the refractive index of PDMS in the IR range, see Table 1. From the visibility of the inference fringes, we infer the transmission coefficient, see Fig. 4(a). In a similar way, we collect the data at different wavelengths. The tunability is achieved by rotating the crystal and selecting the signal photon wavelength by the monochromator. The data is collected with the step of 1 nm for signal photons, which corresponds to the step of 10 nm for idler photons. The obtained transmission spectrum of the PDMS sample is shown in Fig. 4(a). The obtained spectrum is in good agreement with the data for the same sample obtained with a commercial IR spectrophotometer (Shimadzu UV-3600 Plus). The spectral resolution of our technique in the IR is ~2.5 nm at 2200 nm, which is on par with commercially available FTIR systems. It is limited by the resolution of the monochromator for visible photons (~0.25 nm).

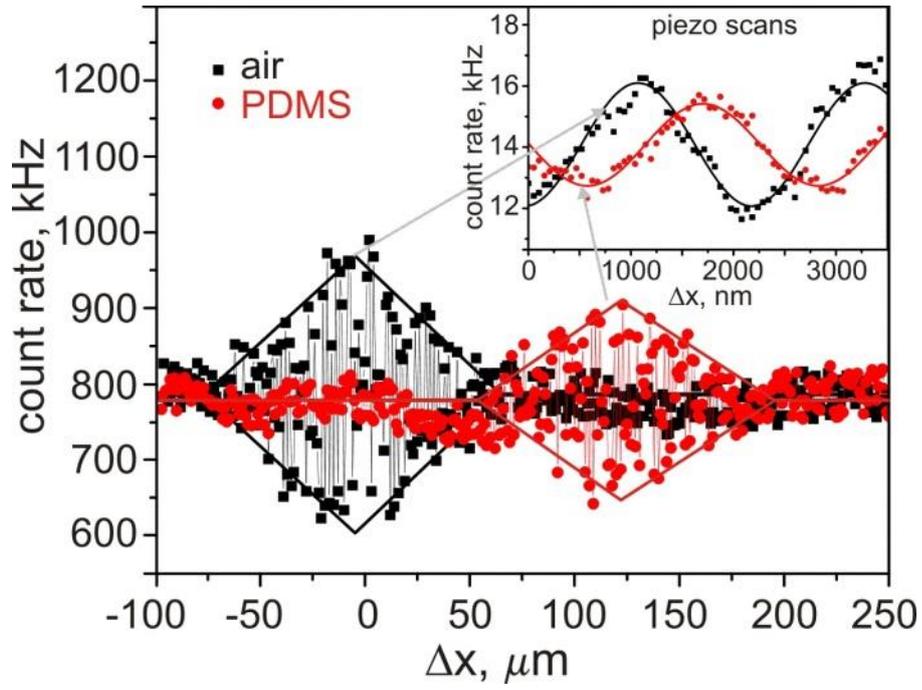

FIG. 3. Dependence of the photocount rate of the APD measured on the position of the mirror $M_i$ (without the monochromator). Black squares show the data with the air, and red circles show the data obtained with a 143 μm thick PDMS sample. Solid lines show a theoretical function given by Eq. (12b) plotted with measured experimental parameters of the interference visibility and the SPDC coherence length. The inset shows the interference fringes observed by fine translation of the mirror $M_i$ using a piezo actuator (with the monochromator). Solid lines show data fitted by cosine-function according to Eq. (12a) with $R^2=0.95$. The absorption coefficient is inferred from the amplitude of the fitting function.

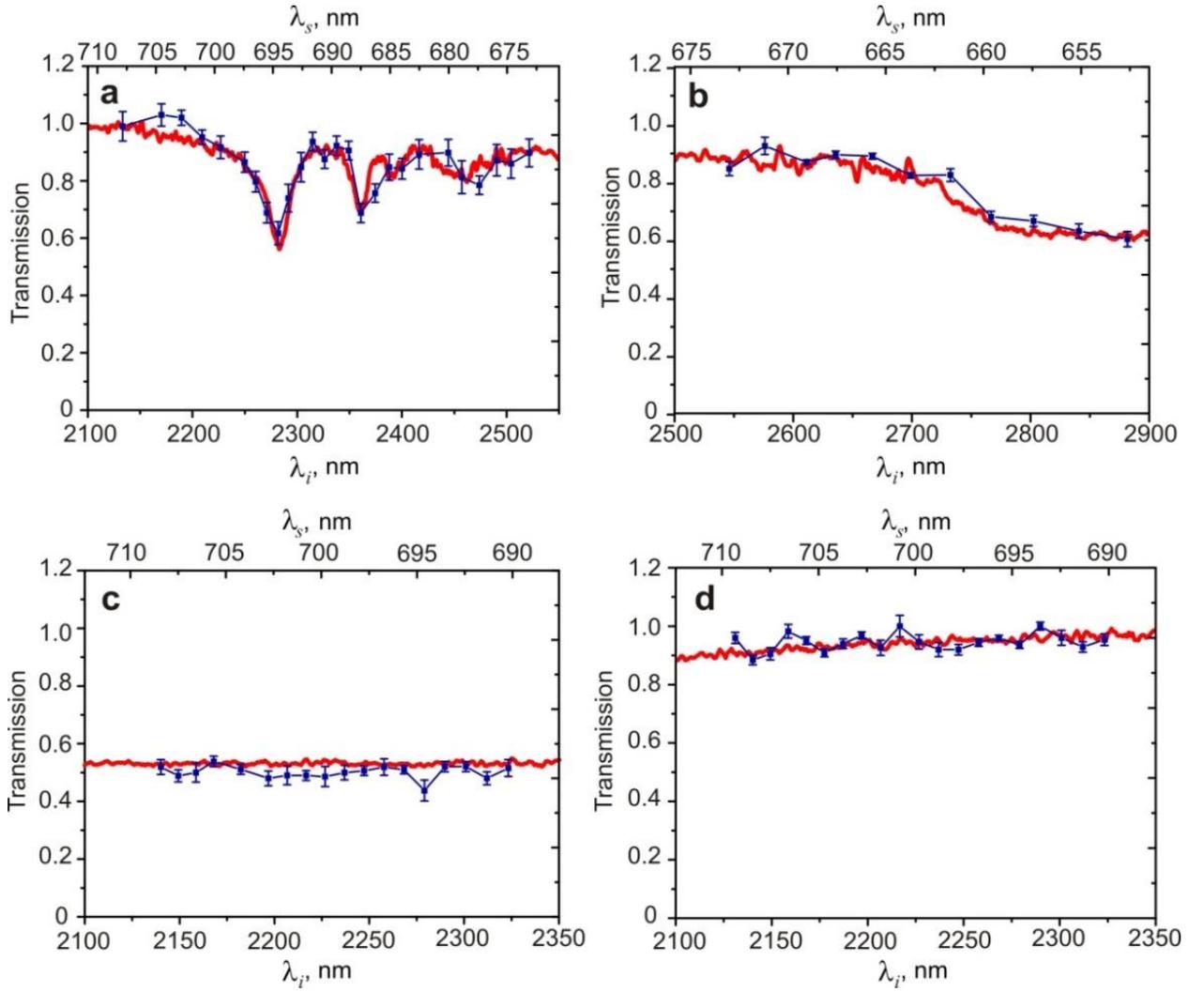

FIG. 4. Transmission spectra of a 143 μm thick PDMS sample (a), a 1 mm thick BK7 sample (b), a 1 mm thick silicon sample without AR coating (c), and a 1 mm thick silicon sample with AR-coating (d). The blue squares show the experimental results obtained by our method; the red curves show the data obtained with a commercial IR spectrophotometer. The top axis indicates the actual measurement wavelength of signal photons and the bottom axis indicates wavelength of idler photons calculated using Eq. (1). Error bars in (a)–(d) show ± s.d.

The spectral range of our technique is defined by the transparency range of the LiNbO3 crystal (0.4-5 μm). We can further extend the range to mid- and far IR range by the appropriate choice of nonlinear crystals. For instance, using $AgGaS_2$ (AGS) crystal will allow extending the transparency range to 15 μm. Since the SPDC is a routine technique used for generation of THz radiation, we suggest that our technique is also applicable to this range [33].

We apply the same measurement procedure for BK7 glass, and AR coated and non-coated silicon samples (1 mm thickness each). The measured transmission spectra are shown in Figs. 4(b), 4(c), 4(d), respectively. Blue squares correspond to the data obtained by our method, and red lines show the data obtained using the commercial IR spectrophotometer. The two sets of data are in good agreement, which proves the validity of our technique.

A relatively thick sample in the idler arm introduces an additional optical delay, which should be compensated by translation of the mirror in the visible/pump arm. Mirror translation may lead to misalignment of the interferometer since the focus of the spherical mirror in the signal/pump beam is displaced from the SPDC crystal. To recover interference fringes in experiments with thick samples, we introduce a well-characterized glass plate in the signal/pump arm which compensates the optical thickness of the sample.

Data on the refractive indices are summarized in Table I and they are in good agreement with the literature [34-36]. The accuracy of the refractive index measurements is $\Delta n'_i = 0.04$ and of the transmissivity is $\Delta T = 0.03$. Reflection and absorption coefficients are calculated by Eq. (17) using the obtained refractive index and transmission coefficient data, see Table II. According to Eq. (17), the accuracy of the reflection coefficient is $\Delta R = 0.01$ and the accuracy of the absorption coefficient is $\Delta \alpha = 0.06$ cm$^{-1}$.

The accuracy of the measurements of the absorption coefficient is determined by the uncertainty of the measurements of the interference visibility. Due to the high count rate of signal photons (tens of kHz) the relative uncertainty in the measured visibility is small (at the order of 0.4%). This enables accurate and precise measurements of the absorption coefficient even with the relatively low value of the visibility (~15%).

It is interesting to compare the resolution and sensitivity of the current method with the one described in [8, 9]. For both configurations the spectral resolution is similar (~2.5 nm at 2200 nm) since it is defined by the spectral resolution of the monochromator for signal photons. The accuracy of the measurements of the absorption coefficient is slightly better for the current method: ~0.1 cm$^{-1}$ in Ref. [8] and ~0.06 cm$^{-1}$ in the current work.

Light scattering in samples limits the performance of spectroscopic and interferometric instruments. In comparison to the earlier implementation of the nonlinear spectroscopy [8, 9], where all the three photons traveled through the sample, the presented technique does not require the interaction of visible light (pump and signal photons) with the sample. Since the major contribution to the scattering occurs at visible wavelengths, our method would be less affected by the scattering. Thus, by probing the samples with IR light, we can achieve higher sensitivity and larger penetration depth in experiments with highly scattering samples.

TABLE I. Values of the refractive index and transmittance (mean ± s.d.) of the samples calculated from the measurements shown in Figs. 3, 4 and using Eqs. (15), (16)

| Sample (thickness, μm) | wavelength, nm | Refractive index, $n'_i$ | | Transmittance, $T_i$ | |
|---|---|---|---|---|---|
| | | our method | database | our method | conventional method |
| PDMS (143±2) | 2200 | 1.41±0.02 | 1.41 | 0.93±0.02 | 0.93 |
| BK7 (980±10) | 2600 | 1.51±0.03 | 1.48 | 0.87±0.02 | 0.88 |
| Si / w/o AR (1030±10) | 2200 | 3.50±0.07 | 3.45 | 0.49±0.04 | 0.52 |
| Si/ w AR (1050±10) | | | | 0.95±0.02 | 0.94 |

TABLE II. Calculated reflectance and absorption coefficients (mean ± s.d.) of the samples using Eqs. (17)

| Sample | wavelength, nm | Reflectance, $R_i$ | | Absorption, $\alpha_i$ cm$^{-1}$ | |
|---|---|---|---|---|---|
| | | our method | database | our method | conventional method |
| PDMS | 2200 | 0.030±0.003 | 0.029 | 0.67±0.06 | 0.69 |
| BK7 | 2600 | 0.041±0.004 | 0.038 | 0.57±0.06 | 0.51 |
| Si /w/o AR | 2200 | 0.31±0.01 | 0.30 | 0.00±0.05 | 0.00 |
| Si/ w AR | | 0.025±0.005 | 0.30 | 0.00±0.01 | 0.00 |

## IV. CONCLUSIONS

In conclusion, we have developed the method for measuring optical constants in the mid-IR range through measurements of the interference of visible photons. The method is based on the nonlinear interference of correlated photons produced via SPDC. Unlike the conventional up-conversion spectroscopy, our technique essentially exploits vacuum fluctuations of the field, which are mixed with the pump laser in a nonlinear crystal. Hence, our approach allows using a simple laser system, consisting of a single visible range cw-laser with fixed wavelength.

Our method allows the calculation of refractive index, transmission, reflection and absorption of samples in a broad spectral range with good accuracy and spectral resolution. This represents a clear advantage over traditional up-conversion spectroscopy which is a transmission-probe technique. Realization of the technique in the high-gain regime may potentially lead to a higher signal-to-noise ratio and the possibility to use conventional photodetectors. However, special precautions are required for achieving high visibility in this regime [10, 37].

Our technique is demonstrated to be applicable to samples opaque in the visible range, which is highly relevant to characterization of optical materials for applications in IR-photonics, telecom, and material analysis. We achieve this by separating idler and signal photons in the interferometer. We also eliminate the requirement of prior information about the sample properties at the visible range, which is a critical limitation of earlier work. The obtained results are in good agreement with data obtained using conventional IR spectroscopy technique and literature data [34-36]. The accuracy of the measurement of the refractive index ($10^{-2}$) is on par with state-of-art refractometers and ellipsometers.

We believe that the work paves the way for further practical adoption of the method in the field of IR and THz optical characterization, without resorting to IR and THz grade detectors and sources.

**ACKNOLEGMENTS**

The work is supported by DSI core funding and by the "Quantum technologies for engineering" (QTE) program of A*STAR. The authors thank D.H. Zhang, G. Vienne, R. Bakker and S. Kulik for stimulating discussions. A.P. would like to acknowledge the support of the SINGA PhD fellowship.